\documentclass[aps,pre,print,tightenlines,twoside,twocolumn,floatfix,floats,amsmath,amssymb,superscriptaddress,altaffillsymbol,nofootinbib]{revtex4-2}
\usepackage{graphicx}
\usepackage[utf8]{inputenx}
\usepackage[T1]{fontenc}
\usepackage[final]{microtype}
\usepackage[x11names]{xcolor}
\usepackage{epstopdf}
\usepackage{dcolumn}
\usepackage{bm}
\usepackage{hyperref}
\hypersetup{
	colorlinks,
	linkcolor={blue!85!black},
	citecolor={blue!85!black},
	urlcolor={blue!85!black},
	pdfauthor = {V. Dzanic, C. S. From, and E. Sauret},
	pdftitle = {Conserving elastic turbulence numerically using artificial diffusivity}
	}
\DeclareMathAlphabet{\mathsfit}{T1}{\sfdefault}{\mddefault}{\sldefault}
\SetMathAlphabet{\mathsfit}{bold}{T1}{\sfdefault}{\bfdefault}{\sldefault}
\providecommand\bnabla{\boldsymbol{\nabla}}
\providecommand\bcdot{\boldsymbol{\cdot}}
\DeclareMathAlphabet\mathsfbi            {OT1}{cmss}{m}{sl}
\IfFileExists{t1phv.fd}
{\DeclareTextFontCommand\textsfbi{\usefont{T1}{phv}{b}{it}}
	\DeclareMathAlphabet\mathsfbi            {T1}{phv}{b}{it}
}{}
\IfFileExists{ot1phv.fd}
{\DeclareTextFontCommand\textsfbi{\usefont{OT1}{phv}{b}{it}}
	\DeclareMathAlphabet\mathsfbi            {OT1}{phv}{b}{it}
}{}
\begin{document}

\title{Conserving elastic turbulence numerically using artificial diffusivity}

\author{V. Dzanic}
\affiliation{School of Mechanical, Medical, and Process Engineering, Faculty of Engineering, Queensland University of Technology, QLD 4001, Australia}
\author{C. S. From}
\email{christopher.from@manchester.ac.uk}
\affiliation{Department of Chemical Engineering, University of Manchester, Manchester M13 9PL, UK}
\author{E. Sauret} \email{emilie.sauret@qut.edu.au}
\affiliation{School of Mechanical, Medical, and Process Engineering, Faculty of Engineering, Queensland University of Technology, QLD 4001, Australia}

\begin{abstract}
%[Limit: ≤ 600 characters (Currently 504)] 
% We demonstrate the applicability of a new artificial diffusivity scheme to simulate elastic turbulence using both the popular cellular forcing and four-roll mill cases. Through a comparative investigation, we show that the new scheme is capable of overcoming the unphysical artifacts that have long-plagued previous artificial diffusivity schemes, allowing the capability to retain key features of elastic turbulence, such as chaotic flow fluctuations characterized by a fairly steep power-law spectrum.  
To simulate elastic turbulence, where viscoelasticity dominates, numerical solvers introduce an artificial stress diffusivity term to handle the steep polymer stress gradients that ensue. This has recently been shown [Gupta \& Vincenzi, \textit{J. Fluid Mech.} \textbf{870}, 405-418 (2019); Dzanic \textit{et al., J. Fluid Mech.} \textbf{937}, A31 (2022)] to introduce unphysical artifacts with a detrimental impact on simulations. In this Letter, we propose that artificial diffusion is limited to regions where stress gradients are steep instead of seeking the zero-diffusivity limit. Through the cellular forcing and four-roll mill problem, we demonstrate that this modified artificial diffusivity is devoid of unphysical artifacts, allowing all features of elastic turbulence to be retained. Results are found to conform with direct simulations
, reducing the impact of artificial diffusivity from a qualitative scale to a quantitative scale
while only requiring a fraction of the numerical resolution.
\end{abstract}

\maketitle

%Non-Newtonian fluids exhibit interesting non-linear material properties, which offer a range of very exciting practical benefits. Viscoelastic fluids, a subclass of Non-Newtonian fluids, are no exception to this.  
It is well-known that fluid flow with the addition of polymer molecules to a solvent (i.e., viscoelastic fluids) in the absence of inertial instabilities $Re= U \ell/ \nu \leq 1$ generates an anisotropic elastic stress contribution that transitions the flow to a novel chaotic regime, known as elastic turbulence (ET) \cite{Groisman_2001, Groisman_2004,Steinberg_REVIEW}. 
ET is purely driven by viscoelastic instabilities, where the viscous to elastic effects are measured by the Weissenberg number $Wi=\tau U/\ell\gg1$, and viscoelastic and inertial effects characterized by the elasticity number $El\equiv Wi/Re = \tau \nu /\ell^2\gg1$.
%, as per most upto date understanding. Re alone is not enough*]
Here, $\tau$ is the longest polymer relaxation time, $\ell$ characteristic length scale, $\nu$ is the total kinematic viscosity, and $U$ is the average velocity.
This purely elastic instability shares a lot of features with traditional inertial turbulence, namely, (i) increased flow resistance, (ii) enhanced mixing, (iii) random flow fluctuations characterized by a broadband spectrum of spatial and temporal frequencies \citep{Steinberg_REVIEW}. 
In an effort to better understand the role of viscoelastic instabilities in these novel chaotic flow regimes, a plethora of numerical studies have been conducted \cite{Poole_2007, Berti_2008, Grilli_2013, plan_gupta_vincenzi_gibbon_2017,gupta_vincenzi_2019,Alves_Rev}. 
% To numerically simulate ET, the Navier-Stokes equations are coupled with convective Maxwell model ...
The majority of which, involve resolving the hydrodynamic field through the incompressible Navier-Stokes equations,
\begin{equation}\label{eq:1}
    \bnabla \bcdot \bm{u} = 0, ~~ \frac{D\mathbf{u}}{D t}=-\bnabla P + \nu_s\mathbf{\Delta}\mathbf{u}+\bnabla \bcdot \bm{\sigma} + \mathbf{F},
\end{equation}
coupled with the the polymer stress tensor $\bm{\sigma}=f(r)\frac{\nu_p}{\tau}\left(\mathsfbi{C}-\mathsfbi{I}\right)$, described by a space-time dependent conformation tensor ($\mathsfbi{C}$) constitutive equation,
\begin{equation}\label{eq:2}
    \frac{D\mathsfbi{C}}{D t} = \mathsfbi{C}\bcdot\left(\bnabla\mathbf{u}\right)+\left(\bnabla\mathbf{u}\right)^{\mathsf{T}} \bcdot \mathsfbi{C}-\frac{f(r)}{\tau}\left(\mathsfbi{C}-\mathsfbi{I}\right),
\end{equation}
where the function $f(r)$ allows for various constitutive polymer models to be used. $\mathsfbi{I}$ is the identity tensor, $\mathbf{u}$ is the velocity field, $\mathbf{F}$ is the external force, $\nu_s$ and $\nu_p$ are the solvent and polymer kinematic viscosity, respectively.

Numerical simulations of turbulent viscoelastic fluid flows is, however, far from trivial. By definition, the conformation tensor is a symmetric positive-definite (SPD) tensor $\mathsfbi{C} \gneq 0$, where conserving this property is important to prevent the rapid growth of Hadamard instabilities \cite{SURESHKUMAR199553}.
To overcome stability issues, numerical methods apply decomposition techniques in Eq.~(\ref{eq:2}) to conserve the SPD properties of $\mathsfbi{C}$ by construction \cite{VAITHIANATHAN20031, Fattal2005TimedependentSO}.
However, these specialized techniques alone cannot control the steep polymer stress gradients that ensue at high elastic effects, captured by the infamous high-$Wi$ number problem, owing to the inherent hyperbolic nature and lack of numerical regularization terms in  Eq.~(\ref{eq:2}).
In turn, solvers introduce an additional global artificial diffusivity (GAD) term in the Laplacian form, $\mathsfbi{D} = k\mathbf{\Delta}\mathsfbi{C}$, into Eq.~(\ref{eq:2})  \cite{gupta_vincenzi_2019,Alves_Rev,LEE_AD}, where the level of diffusivity $k$ is characterized by the Schmidt number $Sc=\nu_s/k$. 
GAD effectively converts the constitutive polymer model to a parabolic form by smoothing steep polymer stresses over large regions of the flow. 
Indeed, polymer-stress diffusion is physically present at $Sc\sim 10^6$; however, to achieve numerical stability requires $Sc \lesssim 10^3$ \cite{VAITHIANATHAN20063, gupta_vincenzi_2019}. 
% as observed experimentally \cite{ElKareh1989ExistenceOS}, however, is at least three orders of magnitude smaller than the $k$ required to achieve numerical stability \cite{VAITHIANATHAN20063,gupta_vincenzi_2019}. 
The unphysically large $k$ values required to achieve such $Sc$ numbers with GAD are known to promote laminarization for elasto-inertial turbulence (EIT) ($Re\gg10$) \cite{Sid_2018}, suppressing the elastic instabilities that are necessary to promote turbulence. %and leading to notable qualitative differences.
% LAD part was here originally.
By combining SPD-conserving solvers with high-order, shock-capturing schemes have enabled simulations of EIT without the need for GAD \cite{shekar_mcmullen_mckeon_graham_2020}.
This same approach was recently used to simulate ET at $Sc=\infty$, where direct comparison against GAD with $Sc=10^3$ revealed the dramatic effect on the large-scale properties of the flow, suppressing chaotic fluctuations \cite{gupta_vincenzi_2019}. %However, \cite{gupta_vincenzi_2019} is also an exemplary indication on the unfeasible resolution required for problems where $El\gg1$.
In the authors' recent work \cite{vedad_jfm}, it was further shown that the global effect of GAD is so large that a single unit cell in fully-periodic problems does not conserve unicity, 
%(i.e., a loss of the underlying structure of the problem), 
leading to numerical artifacts represented by qualitative anomalies, thus rendering the problem unphysical. Conserving unicity allows features of ET to be %partially 
recovered at $Sc=10^3$, with chaotic fluctuations increasing as $Sc>10^3$, whereas ET is suppressed at $Sc\sim10^2$, demonstrating the detrimental effect of GAD. 
Indeed, in the absence of artificial diffusivity, an exact physical representation can be obtained directly from Eq.~(\ref{eq:2}), however, for problems where $El\gg1$, the steep polymer stress gradients that develop require significant spatial and temporal resolution, and hence computational costs, to overcome the numerical stability issues that ensue \cite{plan_gupta_vincenzi_gibbon_2017,gupta_vincenzi_2019,vedad_jfm}. This is in particular problematic given that simulations of ET, in general, are restricted by computational cost due to the inherently small time step required compared to their Newtonian counterpart  \cite{Press_2007, Alves_Rev}.
%\textbf{[Something on how our JFM paper and Gupta's paper raise the questions as to how ET can be reliably simulated while retaining all features. E.g., it is not possible to simply decrease $k$ without running into stability issues and its not feasible to run without it at large grid sizes.]}

In this Letter,
a simple yet elegant approach to include artificially diffusivity while retaining all features of ET is proposed. We demonstrate the ability to avoid the unphysical  artifacts that have plagued the GAD scheme while still offering the same level of numerical benefits, particularly in terms of convenience and stability. 

Earlier works on turbulent polymer drag reduction have seen success using a local artificial diffusivity (LAD) scheme in which $\mathsfbi{D}(\bm{x},t) = k\mathbf{\Delta}\mathsfbi{C}(\bm{x},t)$ is applied in regions of the computational domain $\bm{x}$ where $\mathsfbi{C}\leq0$ \cite{Min_2001,Dubief_2005}. However, LAD does not guarantee SPD and is inherently not applicable to modern, SPD-conserving solvers \cite{VAITHIANATHAN20031, Fattal2005TimedependentSO,Alves_Rev}.  Motivated by the LAD scheme, as well as the stability rationale, i.e., the singular purpose of artificial diffusivity is for numerical regularity, we seek a suitable approach with which $k$ is not decreased but instead restricted where it is applied to minimize the potential consequent global effects. %and previous sub-grid scale models for compressible turbulence (i.e., large-eddy simulation (LES) \citep{Javier_LES}),
%large-scale effects
%, as well as our previous work \citep{vedad_jfm}, which showed that as $Sc\to\infty$ features of ET becomes more apparent, 
To this aim, we propose a modified artificial diffusivity (MAD) scheme such that $\mathsfbi{D}(\bm{x},t) = \kappa(\bm{x},t)\mathbf{\Delta}\mathsfbi{C}(\bm{x},t)$ with $\max_{\bm{x}}[\kappa(\bm{x},t)]=k~\forall t$, where $\bm{x}$ is the domain, which is directly applicable to all modern, SPD-conserving, numerical solvers. An intuitive way to apply $\kappa$ is to consider the characteristic feature of ET, that the stress gradients are sharp and localised. By restricting artificial diffusivity within those regions to a maximum of $k=\nu_s/Sc$, we allow for the required numerical regularity to be retained while minimizing global smearing where regularization is not needed. Leveraging  this idea, multiple possible variations for $\kappa(\bm{x},t)$ naturally emerge (as discussed in the Supplementary Material). To illustrate, we consider the Laplacian form,
\begin{equation}\label{eq:3}
\mathsfbi{D}(\bm{x},t) = 
%\left(
\kappa(\bm{x},t)
% \frac{k \mathcal{Q}(\bm{x},t)}{\mathcal{Q}_{max}(t)}
% \right) 
\mathbf{\Delta}\mathsfbi{C}(\bm{x},t),
~\kappa(\bm{x},t) = \frac{k \mathcal{Q}(\bm{x},t)}{\mathcal{Q}_{max}(t)},
\end{equation}
where $\mathcal{Q}$ is the sum of the magnitude of polymer stress component gradients, i.e.,
$$ 
\mathcal{Q} (\bm{x},t) = \sum_{i,j}^{D} \sqrt{\sum_{q}^{D} \Big[\nabla_{q}\mathsfit{C}_{ij}(\bm{x},t)\Big]^2}, 
$$
for a $D$ dimensional domain $\bm{x}$ and $ \mathcal{Q}_{max} (t) = \max_{\bm{x}} [\mathcal{Q} (\bm{x},t)]$ is a normalization factor. 
% $$
% \mathcal{Q} (\bm{x},t) = \parallel\bm{\nabla\sigma}(\bm{x},t)\parallel
% $$
In turn, in MAD (\ref{eq:3}) the artificial diffusivity is scaled linearly with stress gradients to a maximum of $k$, i.e., 
$\lim_{\mathcal{Q}\rightarrow\mathcal{Q}_{max}} \mathsfbi{D} = k\mathbf{\Delta}\mathsfbi{C}$. 
%\textcolor{blue}{\textbf{[I think we can probably delete this.]} The authors note that the expression for $\kappa$ (\ref{eq:3}) is one of several variations of MAD, with other alternatives discussed in the Supplementary Material.}   
%

%\textbf{[TODO--> Revise]}
%In principle, the MAD scheme is applied \textcolor{red}{globally} in the same manner as the GAD scheme, while echoing certain features from the LAD scheme. More specifically, it can be seen from (\ref{eq:3}) \textcolor{red}{that although applied globally}, artificial diffusivity is mostly applied within the regions of high polymer stress gradients, whereas $k\to0$ for regions where $\mathcal{Q}\ll \mathcal{Q}_{max}$.
%////

\begin{figure}[t!]
	\centering
	\includegraphics[width=1\columnwidth]{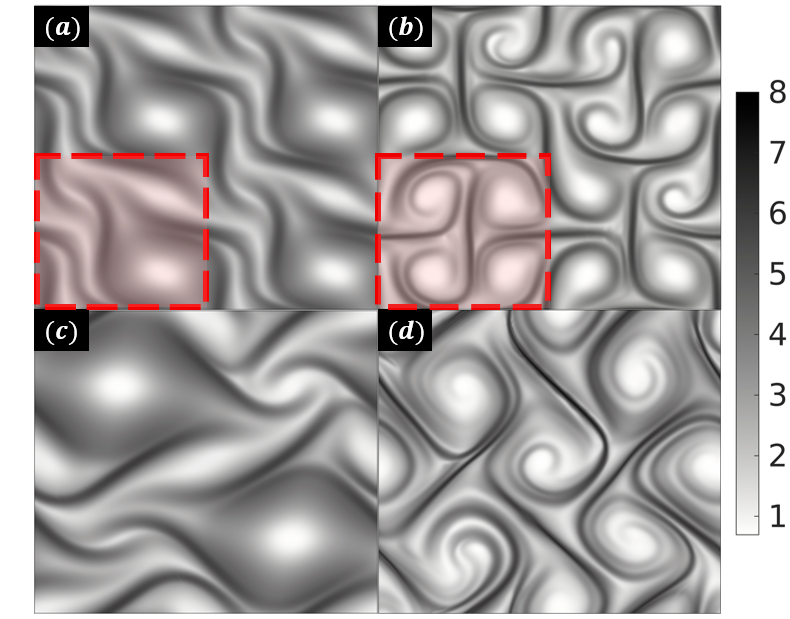}%scale=0.4%width=0.95\textwidth
	\caption{Representative snapshots of $\ln($tr$\mathsfbi{C})$ for FRM (a,~b) with $n=2$ and CF (c,~d) using GAD (a,~c) and MAD (b,~d). The red borders in (a,~b) illustrate a single unit cell for FRM. For both cases, simulations with GAD (a,~c) result in a loss of symmetry spawned from artificial diffusivity, whereas with MAD (b,~d) they are largely constrained to the background forcing symmetry.}
	\label{figure_1}
\end{figure}
\begin{figure}[b!]
	\centering
	\includegraphics[width=1\columnwidth]{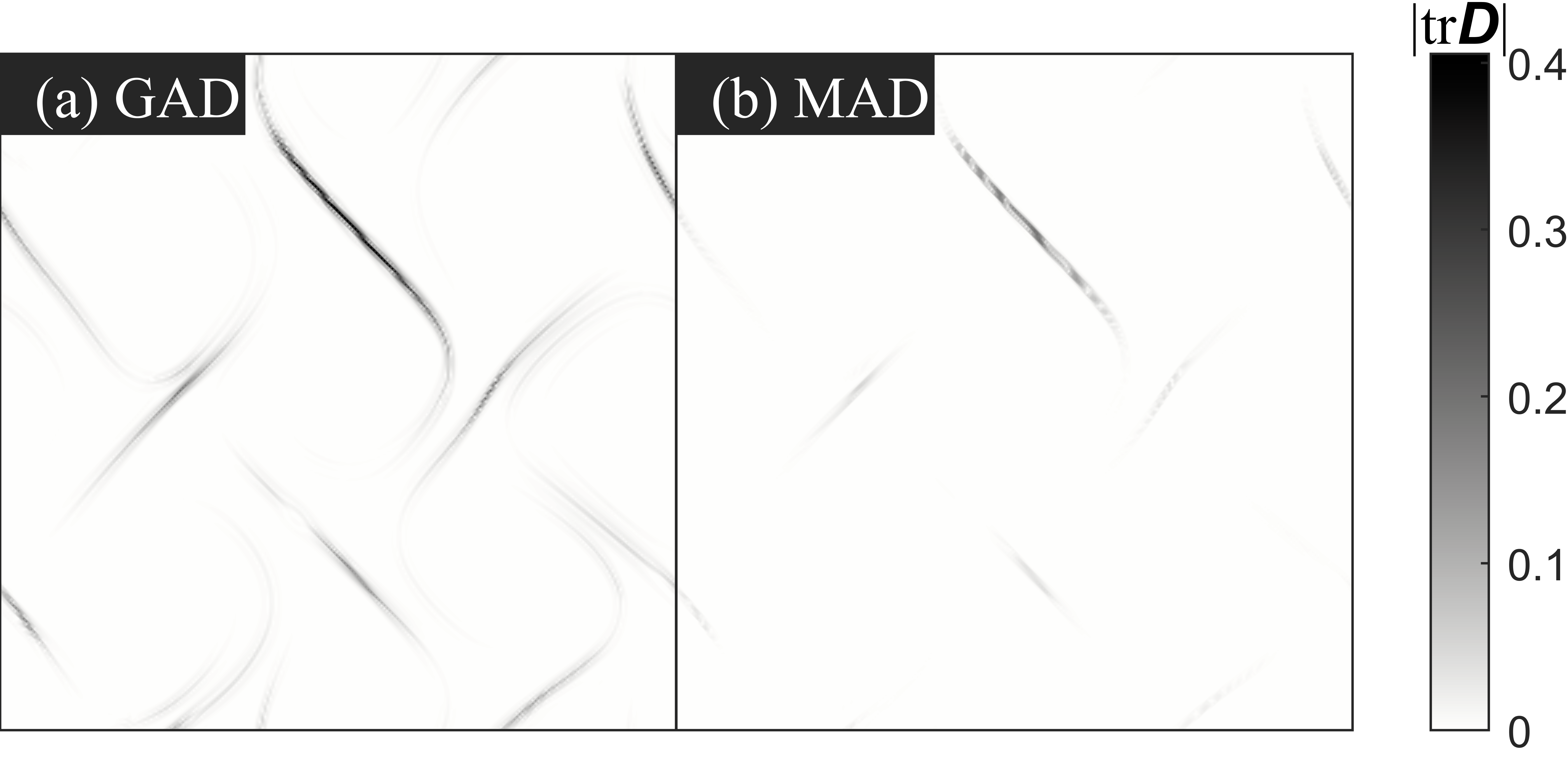}
	\caption{The $|$tr$\mathsfbi{D}|$ obtained with (a) GAD $\mathsfbi{D} = k\mathbf{\Delta}\mathsfbi{C}$ and (b) MAD [Eq.~(\ref{eq:3})] using $\mathsfbi{C}$ from the simulation of CF with MAD in Fig.~\ref{figure_1} (d). With MAD, $\mathsfbi{D}$ is largely concentrated at steep stress gradients where $\forall\kappa\gneq0,~ Sc = 10^3 \sim 10^8$.}
	%, whereas with GAD $Sc$ is constant $10^3$.}
	\label{figure_2}
\end{figure}
\begin{figure*}[t!]
	\centering
	\includegraphics[scale=0.74]{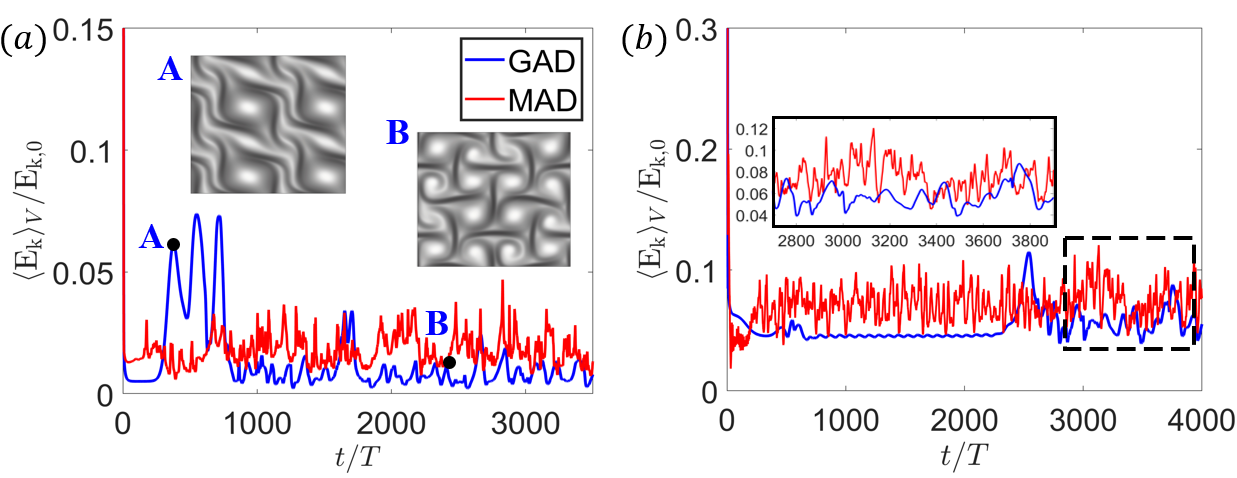}%width=0.95\textwidth
	\caption{Time series of the dimensionless mean kinetic energy $\langle E_{k}\rangle_{V}/E_{k,0}$ for the (a) FRM and (b) CF using the GAD (blue) and MAD (red) schemes. Note, insets provided in (a) correspond to the snapshots of $\ln($tr$\mathsfbi{C})$ at locations A ($t=400T$) and B ($t=2500T$) for the GAD scheme, illustrating the different transition points. The inset provided in (b) is a zoom-in corresponding to the late-time dynamics in the dashed box region.}
	\label{figure_3}
\end{figure*}

To demonstrate the suitability of the MAD scheme while further demonstrating the severe limitations of the GAD scheme for simulations of ET, we apply both MAD and GAD to simulate two popular ET cases, namely, the four-roll mill (FRM) problem \cite{T_S_2007,T_S_2009,THOMASES20111221} and the cellular forcing (CF) scheme \cite{plan_gupta_vincenzi_gibbon_2017,gupta_vincenzi_2019}. In recent investigations of ET, simulations of FRM and CF using GAD were shown to depict various numerical artifacts that have been investigated extensively in both \cite{vedad_jfm} and \cite{gupta_vincenzi_2019}, respectively, which presents strict test criteria for MAD.
Both FRM and CF are numerical recreations of popular viscoelastic experimental cases \cite{Cardoso_1994, Rothstein1999PersistentPI, LIU_2012}, and are solved in a 2D domain $\bm{x}=[0,~n\times 2\pi]^2$ with double periodic boundary conditions (PBCs) where a single unit cell is $\left[0,2\pi\right]^2$. Here, $n$ is the level of periodicity, for which $n>1$ results in $n^2$ unit cells to be solved. 
The experimental effect of rollers, which create an elongational flow regime, are mimicked through a constant external forcing, which for the FRM problem is given by 
\begin{equation}\label{eq:FRM}
    \mathbf{F} (\bm{x}) = F_0\bigl(2\sin\left(K{x}\right)\cos\left(K{y}\right),-2\cos\left(K{x}\right)\sin\left(K{y}\right)\bigr),
\end{equation} 
and for the CF forcing scheme, 
\begin{equation}\label{eq:CF}
    \mathbf{F} (\bm{x}) = F_0\left(-\sin\left(K{y}\right),\sin\left(K{x}\right)\right),
\end{equation}
in which the forcing amplitude is $F_0=U\nu_s K^2$ and $K$ is the spatial frequency (i.e., $\ell = 1/K$), resulting in a turnover time $T = \nu_s K/F_0$. For FRM (\ref{eq:FRM}) $K=1$ and for CF (\ref{eq:CF}) $K=2$. To simulate ET using FRM and CF, a small perturbation $\bm{\delta}$ is added to the initial conformation tensor $\mathsfbi{C}=\mathsfbi{I}+\bm{\delta}$, as originally proposed in \cite{T_S_2009}. Equations~(\ref{eq:1}) and (\ref{eq:2}), with GAD and MAD (\ref{eq:3}), are solved using a SPD-conserving numerical solver comprising of the lattice Boltzmann coupled with a high-order finite-difference scheme (see \cite{DZANIC2022105280}), which was applied in our previous investigation of ET \cite{vedad_jfm}. Wherein, the lattice Boltzmann method inherently permits exact advection for the hydrodynamic field, and is thus devoid of numerical diffusion in Eq.~(\ref{eq:1}). To directly resolve Eq.~(\ref{eq:2}),  the polymer solver treats the advection term according to the high-resolution Kurganov-Tadmor scheme \cite{KT_SCHEME}, while a fourth-order Runge-Kutta scheme is applied for the temporal evolution. 
Spatial gradients in the artificial diffusivity terms, including (\ref{eq:3}), are solved using a second-order central difference scheme, whereby any consequent numerical diffusion is comparatively much lower than the added levels of artificial diffusivity, and thus, has been shown to have a negligible effect on ET \cite{gupta_vincenzi_2019, vedad_jfm}, retrieving results in direct agreement with previous spectral studies of viscoelastic instabilities \cite{T_S_2007, T_S_2009,THOMASES20111221,DZANIC2022105280}. We realistically capture the main physical behavior of polymers using the FENE-P constitutive model, $f(r) = \left(L^2 - 2\right)/\left(L^2 - r\right)$ in Eq.~(\ref{eq:2}) with $r = \mathrm{tr}\mathsfbi{C}$ and $L^2 = 2.5\times10^3$, which imposes a maximum finite extensibility $L > r$ \cite{PETERLIN1961257}.  
In all simulations, dimensionless groups are set following previous numerical investigations including, $Wi=10$, artificial diffusivity $Sc=10^3$ (as done in  \citep{T_S_2009, THOMASES20111221, gupta_vincenzi_2019}), and set $Re=1$ (as done in \citep{Berti_2008, plan_gupta_vincenzi_gibbon_2017,vedad_jfm}) below the critical value at which inertial instabilities arise $Re_c=\sqrt{2}$ \citep{Gotoh1984InstabilityOA}.
The polymer concentration, characterized by $\beta=\nu_p/\nu_s$, is set to $\beta=0.5$ and $\beta=0.4$ in order to match previous investigations for FRM \cite{T_S_2009,THOMASES20111221} and CF\citep{gupta_vincenzi_2019}, respectively. 

Based on our previous investigation \citep{vedad_jfm}, GAD with $n=1$ will result in unphysical artifacts arising from insufficient periodicity, characterized by qualitative anomalies which contaminate the base flow, thus preventing the accurate simulation of ET. For CF, we set $n=1$ to compare against results from \cite{gupta_vincenzi_2019}, which were solved on a $N^2=1024^2$ grid for both $Sc=10^3$ (using GAD) and $Sc=\infty$. In addition, this is useful for testing whether unicity can be retained. For FRM, where the external forcing is larger, we set $n=2$ to conserve unicity allowing numerical artifacts arising \textit{purely} from artificial diffusivity to be studied.
    % ... This is particularly ... since in FRM the external forcing is larger comparitively larger than CF, which as noted by Gupta enabled direct simulation of CF  N^2=1024^2 --Ugh, not sure with this could work...
All simulations are purposely conducted with $(n\times N)^2=256^2$ grid points, where each unit cell has $N^2$ grid points, i.e., the resolution $2\pi/N$ for FRM is half of that used for CF. This is a notable reduction compared to previous studies \citep{plan_gupta_vincenzi_gibbon_2017,gupta_vincenzi_2019} to highlight the ability to retain numerical robustness with the proposed approach of modified artificial diffusivity. %\textbf{We confirmed that the major findings of this investigation were not grid-dependent.} \textbf{\textcolor{red}{[Where???]}}
% In recent numerical simulations of ET, it was observed that the GAD scheme can give rise to unphysical numerical artifacts, whereby FRM \cite{vedad_jfm} and CF \cite{gupta_vincenzi_2019} are largely characterized by a loss of background forcing symmetry [i.e., Eqs.~(\ref{eq:FRM}) and (\ref{eq:CF})] and the generation of dominant vortices. 
 
\begin{figure*}[t!]
	\centering
	\includegraphics[scale=0.47]{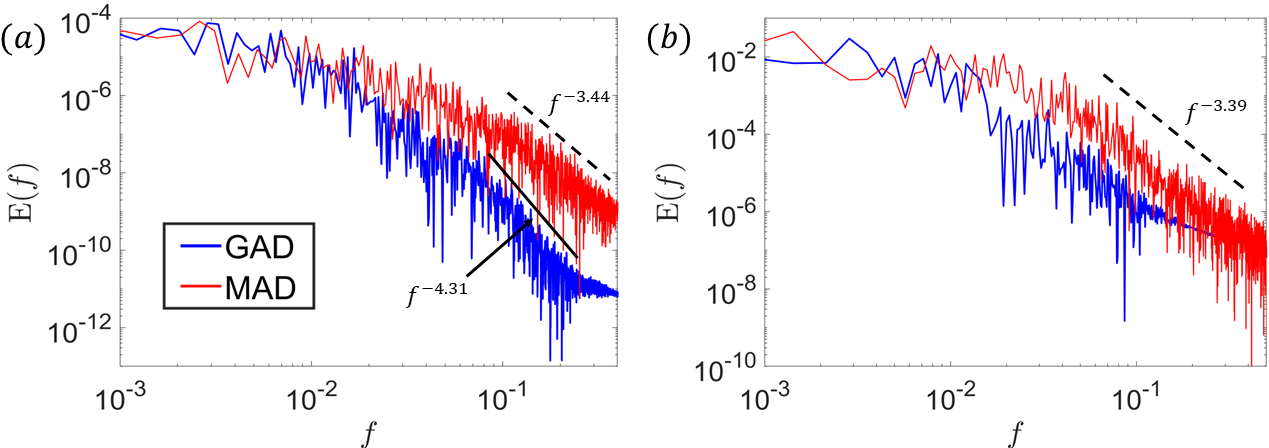}%width=0.95\textwidth
	\caption{The temporal power spectral density $E(f)$ of the dimensionless mean kinetic energy fluctuations for the (a) FRM and (b) CF using the GAD (blue) and MAD (red) schemes. 
	Note, results with MAD follow a power-law, $E\propto f^{-3.44}$ and $E\propto f^{-3.39}$ for the FRM and CF, respectively. With GAD, FRM with $n=2$ follows a steeper power-law $E\propto f^{-4.31}$, while failing to follow a power-law behavior for the CF with $n=1$.
	} 
	\label{figure_4}
\end{figure*}
In Fig.~\ref{figure_1}, contour plots of the polymer field $\ln($tr$\mathsfbi{C})$ using the GAD and MAD are compared for both the FRM [Fig.~\ref{figure_1}~(a) and (b)] and CF [Fig.~\ref{figure_1}~(c) and (d)] cases. The stark differences are immediately clear on a qualitative-scale. More specifically, it can be seen that FRM with GAD leads to a single-leading vortex within each unit cell [note, a unit cell is characterized by four rollers as illustrated by the red borders in Fig.~\ref{figure_1}~(a) and (b)], which appears during the onset of ET. Similarly, CF with GAD eventually transitions from the initial background forcing symmetry into a state with which dynamics are periodic and then into the double-leading vortices in Fig.~\ref{figure_1}~(c), where dynamics become aperiodic. Analogous results are obtained for FRM at $n=1$ using GAD \cite{T_S_2009}.
Ultimately, these are a product of the GAD excessively spreading polymer stresses over large regions of the flow, even within the vortical regions where there should be no polymer stretching, as illustrated in Fig.~\ref{figure_2}~(a). In turn, the polymer field eventually destabilizes and loses the initial forcing symmetry \citep{gupta_vincenzi_2019}. 
The initial breakdown in symmetry during the onset of ET has been shown to be independent of periodicity \citep{vedad_jfm}.
%, highlighting the detrimental global effect of GAD. 
Furthermore, considering the same observations were made for CF with GAD \cite{gupta_vincenzi_2019} but at significantly larger resolution ($N^2=1024^2$) demonstrates the grid-independence of numerical artifacts arising from artificial diffusivity and periodicity.
On the other hand, numerical simulations of FRM and CF with MAD are \textit{devoid} of such numerical artifacts.  
%a loss of symmetry and quickly transitions to a chaotic regime. 
For CF with MAD, despite experiencing momentary losses of symmetry, the polymer stretching is mostly constrained to the background forcing symmetry, in qualitative agreement with the results obtained using direct simulations at $Sc=\infty$ in \citep{gupta_vincenzi_2019} (see animations in the Supplementary Materials). Notably, the momentary losses of symmetry are expected to be eradicated with $n>1$ \citep{vedad_jfm}, nevertheless, we shall see that these have a negligible impact on the ability to simulate ET. Essentially, MAD overcomes numerical artifacts by concentrating $\mathsfbi{D}$ only within the steep polymer stress gradients [Fig.~\ref{figure_2}~(b)], whereas with GAD $\mathsfbi{D}$ is diffused over all gradients  [Fig.~\ref{figure_2}~(a)].

The quantitative results further reflect the stark differences observed for the two artificial diffusivity schemes. The time history response for the dimensionless mean kinetic energy $\langle E_{k}\rangle_{V}/E_{k,0}$, where $\langle\cdot\rangle_{V}$ denotes the spatial average and $E_{k,0}=U^2/2$, is shown for both FRM and CF in Fig.~\ref{figure_3}. Notably, the time series of the spatially-averaged polymer trace is also provided in the Supplementary Material and shows analogous results for both FRM and CF.    
In Fig.~\ref{figure_3}~(a), FRM with GAD initially reaches a steady-state, an artifact due to GAD suppressing intrinsic instabilities, followed by an onset of ET for which there is a rapid loss of symmetry [inset point A in Fig.~\ref{figure_3} (a)], which is characterized by slow oscillations. In our previous study \citep{vedad_jfm}, we showed that a complex interplay exists between periodicity and artificial diffusivity, whereby the classic FRM problem, in which $n=1$, is unable to overcome this initial loss of symmetry, as similarly observed for CF with GAD in Fig.~\ref{figure_3}~(b). Imposing sufficient $n$-levels of periodicity, as done here for FRM with $n=2$, in turn, conserves unicity [e.g., the background forcing symmetry, shown in inset point B in Fig.~\ref{figure_3}~(a)] and enables features of ET to be recovered, such as chaotic flow fluctuations. Notably, defining $n>2$ increases the rate at which ET is recovered after the initial loss of symmetry \cite{vedad_jfm}. 
% MAD part for FRM:
Remarkably, with MAD, the FRM results are devoid of all unphysical artifacts spawned from artificial diffusivity. 
%
% Fig.2 CF Discussion:
When observing the results for CF with GAD in Fig.~\ref{figure_3}~(b), multiple instability modes exist. The first mode occurs within the early stages during an initial partial loss of flow symmetry, which periodically cycles between two different states, reflected by the fully periodic dynamics. This is the first observable numerical artifact caused by artificial diffusivity \cite{gupta_vincenzi_2019} and PBCs with limited periodicity \cite{vedad_jfm}. Qualitatively, this instability mode can be appreciated in the animation provided in the Supplementary Materials. Beyond this state, the flow experiences a complete loss of symmetry, dominated by two vortices [refer to Fig~\ref{figure_1}~(c)] where the dynamics transition into an aperiodic state, as previously reported in \cite{gupta_vincenzi_2019} and analogous results can be observed for FRM with $n=1$  \cite{T_S_2009,THOMASES20111221,DZANIC2022105280}. While the initial loss of symmetry is a consequence of artificial diffusivity, the periodic and aperiodic states are largely due to the limited periodicity, where conserving unicity, as done here for FRM with GAD, prevents such unphysical behavior \citep{vedad_jfm}. 
On the other hand, CF with MAD [Fig.~\ref{figure_3}~(b)] almost instantly transitions into a fully chaotic state, where the heavy fluctuations perturb the cellular vortices, however, the large-scale structure remains largely constrained to the background forcing with momentary losses of symmetry [refer to Fig.~\ref{figure_1}~(d)]. 
In comparing the late-time dynamics within the statistically homogeneous state (approximately $t>500T$, for both FRM and CF), it is noticeable that for CF with MAD,  fluctuations are at a much higher frequency compared to the slow aperiodic state observed for GAD. Similar observations were made in comparing GAD against the direct solution (i.e., $Sc=\infty$) \cite{gupta_vincenzi_2019}. 
Thus, the excessive diffusivity from GAD suppresses the chaotic fluctuations characteristic of the ET regime, a well-reported artifact of artificial diffusivity \citep{gupta_vincenzi_2019,vedad_jfm}.
% Closing statement on Fig.2
High-frequency oscillations with MAD occur almost instantly for both FRM and CF, which is in agreement with the direct simulation of CF at $Sc=\infty$ \cite{gupta_vincenzi_2019} and experimental analog of FRM \citep{LIU_2012}.
% Final note on Fig.2

In Fig.~\ref{figure_4}, we examine the temporal power-law spectrum of
the time signals from Fig.~\ref{figure_3} within the statistically homogeneous regime. A distinctive feature of ET is a fairly steep power-law spectrum of velocity fluctuations \cite{Steinberg_REVIEW,Steinberg_Scaling}.
The fluctuations for FRM with $n=2$ using GAD [Fig.~\ref{figure_4}~(a)] follows a steep power-law with an exponent of $E\propto f^{-4.31}$. This steepness is independent of periodicity for $n\gneq1$, \citep{vedad_jfm} and the results here for FRM with MAD, where $E\propto f^{-3.44}$, confirm this to be attributed to the numerical artifacts.
The aperiodic state observed for CF with GAD [Fig.~\ref{figure_4}~(b)] fails to follow any apparent power-law, as originally observed in \citep{gupta_vincenzi_2019} with significantly greater resolution. 
This is attributed to GAD suppressing the high-wavenumber fluctuations of polymer stresses and is analogously observed for FRM with $n=1$ \cite{vedad_jfm}. Ultimately, with GAD and $n=1$, numerical simulations of FRM and CF are unable to conserve any features of ET.
On the other hand, for both cases, the fluctuations retained in simulations with MAD behave as a power-law, with exponents similar to previous experimental and numerical studies of ET, in which the decay rate varied with the setup, but the exponent was always smaller than $-3$ \cite{Groisman_2001,Groisman_2004,Berti_2008,Canossi_2020,Steinberg_REVIEW, Steinberg_Scaling}.

MAD for turbulent non-Newtonian fluid flow is akin to sub-grid and hyperviscosity models for traditional Newtonian turbulence modeling, e.g., large-eddy simulations \cite{LES_REF}.
Inspired by this, the general form of the diffusion equation employed by traditional turbulence models is considered for MAD, i.e., $\bnabla\bcdot(\kappa\bnabla\mathsfbi{C})$, instead of the simplified Laplacian form (\ref{eq:3}), provided in Supplementary Material. It is found that with the general form diffusion, features of ET are more accurately captured compared with GAD; however, unphysical artifacts persist, unlike with Eq.~(\ref{eq:3}). The underlying cause for differences between $\bnabla\bcdot(\kappa\bnabla\mathsfbi{C})$ and the Laplacian form (\ref{eq:3}) is not currently understood. The authors speculate that perhaps the superior numerical performance of the Laplacian form (\ref{eq:3}) compared to $\bnabla\bcdot(\kappa\bnabla\mathsfbi{C})$ is the likely result of the original treatment of diffusivity [i.e., $\kappa(\bm{x},t)$] originating from the view of diffusion in the Laplacian form. Nevertheless, these comparisons exemplify variations of MAD and the complex interplay between artificial diffusivity and periodicity in ET.
In a similar fashion to traditional turbulence modeling, MAD admittedly introduces additional numerical uncertainties. Firstly, it does not completely omit the polymer field from the presence of artificial diffusivity altogether and, thus, does not completely resolve the small-scale polymer dynamics. Furthermore, linearly scaling the artificial diffusivity with the normalization of the polymer stress gradients introduces an additional source of fluctuation to the polymer field whose large-scale impact is currently not understood. Nevertheless, the strong retention of ET features using MAD is a result of restricting artificial diffusivity within the critical regions of the flow [i.e., steep polymer stress gradients, as shown in Fig.~\ref{figure_2}], thus avoiding the consequent global deviation from the exact polymer representation (i.e., $Sc=\infty$). 
The remarkable feature of MAD (\ref{eq:3}) is the numerical robustness while retaining all features of ET with an absence of numerical artifacts despite using 16 times less collocation points (i.e., a quarter resolution) compared to that required by direct simulations of ET at $Sc=\infty$ \cite{gupta_vincenzi_2019,plan_gupta_vincenzi_gibbon_2017}. 

%Much like the initial development stages of sub-grid scale models for compressible turbulence, the current performance of MAD should not be based on any theoretical understanding but instead its robustness at capturing key characteristics of ET.... 

%Furthermore, MAD does not completely omit the small scale polymer dynamics from artificial diffusivity, it remains a numerical artefact whose impact on the small scale behaviour is not fully understood and under certain conditions can potentially introduce additional unphysical numerical artifacts. %To progress MAD further, future work will involve investigating these uncertainties.

Summarising, a new view on including numerical regularity for simulating elastic turbulence (ET) through a modified artificial diffusivity (MAD) has been proposed. Its applicability is demonstrated by applying the MAD to two stringent numerical experiments.
It is shown that with MAD, all characteristic features of ET can be simulated while overcoming the recently discovered unphysical numerical artifacts of the global artificial diffusivity used traditionally. The impact of artificial diffusion with MAD is effectively reduced from a qualitative scale to a quantitative scale, requiring a numerical resolution that is one order of magnitude smaller compared to that required to numerically simulate ET at $Sc=\infty$. It is difficult to overlook the apparent trend and analogy that artificial diffusivity schemes share with traditional turbulence modeling.
In a similar fashion, the MAD scheme is an additional numerical tool that offers numerical features that could be of paramount importance in progressing towards more complex and computationally expensive ET cases, notably, 3D simulations.

% Summarising, we have demonstrated the applicability of a new modified artificial diffusivity (MAD) scheme for simulating ET. Through application of the scheme to popular stringent test cases, it is shown that the MAD scheme is capable of retaining all characteristic features of ET, while overcoming the unphysical artifacts of global artificial diffusivity, which have long-plagued previous numerical studies of ET.
% The impact of artificial diffusion with MAD is effectively reduced from a qualitative scale to a quantitative scale, requiring a numerical grid that is a magnitude smaller compared to direct solution at $Sc=\infty$.
% It is difficult to overlook the obvious trend and analogy that artificial diffusivity schemes share with traditional turbulence modeling (i.e., RANS, LES, and DNS). In similar fashion, the MAD scheme is an additional numerical tool, which offers numerical features that could be of paramount importance in progressing towards more complex and computationally expensive ET cases, \textbf{notably, 3D simulations}.

The authors acknowledge the High-Performance Computing facilities at QUT. Prof. E. Sauret is the recipient of an Australian Research Council Future Fellowship (FT200100446) funded by the Australian Government. V. Dzanic gratefully acknowledges QUT for support through a Ph.D. scholarship.

\bibliographystyle{apsrev4-1} % Tell bibtex which bibliography style to use
\bibliography{BIB_ET_PAPER}
\end{document}

% Use the \preprint command to place your local institutional report
% number in the upper righthand corner of the title page in preprint mode.
% Multiple \preprint commands are allowed.
% Use the 'preprintnumbers' class option to override journal defaults
% to display numbers if necessary
%\preprint{}
%Title of paper
\title{Conserving elastic turbulence numerically using artificial diffusivity\\
SUPPLEMENTARY MATERIALS}

\author{V. Dzanic}
\affiliation{School of Mechanical, Medical, and Process Engineering, Faculty of Engineering, Queensland University of Technology, QLD 4001, Australia}
\author{C. S. From}
\email{christopher.from@manchester.ac.uk}
\affiliation{Department of Chemical Engineering, University of Manchester, Manchester M13 9PL, UK}
\author{E. Sauret} \email{emilie.sauret@qut.edu.au}
\affiliation{School of Mechanical, Medical, and Process Engineering, Faculty of Engineering, Queensland University of Technology, QLD 4001, Australia}

\maketitle

The supplementary materials consist of additional information and results for an alternative formulation of the modified artificial diffusivity (MAD*) scheme, which is compared against the results for the global artificial diffusivity (GAD) and modified artificial diffusivity (MAD) schemes using the cellular forcing (CF) and the four-roll mill (FRM) cases. Furthermore, additional quantitative results and animations are provided for the polymer field to further support the comparisons made between the GAD and MAD schemes in the main text.

\begin{figure}
	\centering
	\includegraphics[scale=0.68]{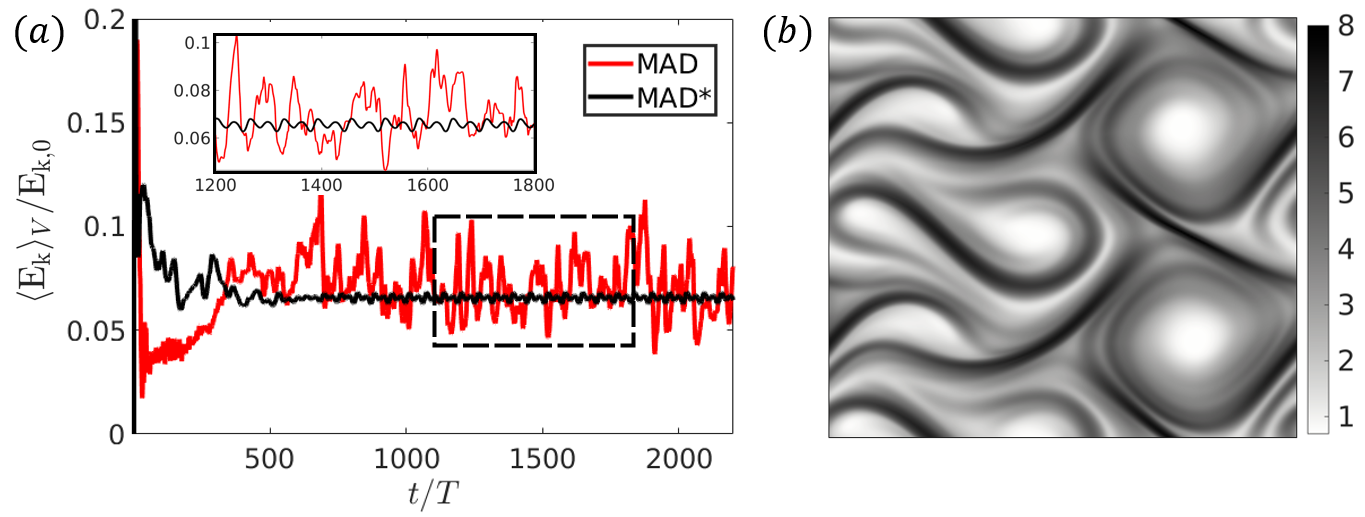}%width=0.95\textwidth
	\caption{(a) Time series of the dimensionless mean kinetic energy $\langle E_{k}\rangle_{V}/E_{k,0}$ for the CF using MAD (red) and MAD* (black) schemes. Note, the inset provided in (a) is a zoom-in section corresponding to the late-time dynamics in the dashed box region. (b) Representative snapshot of $\ln($tr$\mathsfbi{C})$ for CF using MAD*. }
	\label{figure_2_supp}
\end{figure}
\begin{figure}
	\centering
	\includegraphics[scale=0.5]{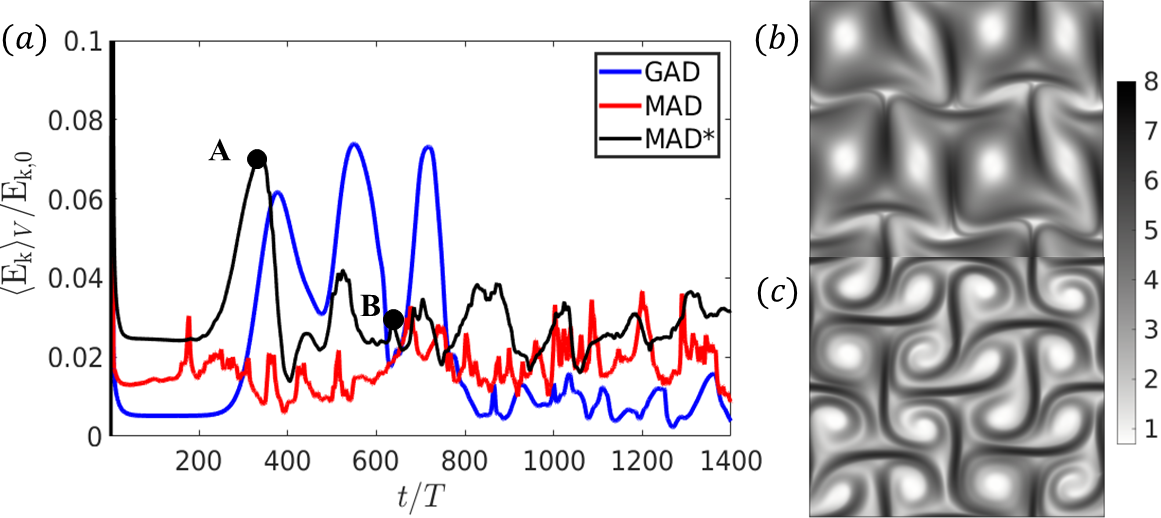}%width=0.95\textwidth
	\caption{(a) Time series of the dimensionless mean kinetic energy $\langle E_{k}\rangle_{V}/E_{k,0}$ for the FRM using GAD (blue), MAD (red), and MAD* (black) schemes.  (b) and (c) Representative snapshots of $\ln($tr$\mathsfbi{C})$ for FRM using MAD* at location A ($t=390T$) and location B ($t=650T$), respectively. }
	\label{figure_3_supp}
\end{figure}

\section{ALTERNATIVE MAD SCHEME FORMULATION}
The fundamental concept of MAD is to have a spatially continuous diffusion coefficient $\kappa$ that is concentrated within regions where the stress gradients are sharp, such that $\max_{\bm{x}}[\kappa(\bm{x},t)]=\nu_s/Sc~\forall t$, which are the regions where regularization is critically required. Naturally, there are many possible approaches to formulate $\kappa(\bm{x},t)$, and in the main text [refer to Eq.~(3)], we illustrate this idea with a simple formulation for $\kappa$, considering the following Laplacian form,
\begin{equation}\label{eq:1_sup}
\mathsfbi{D}(\bm{x},t) = 
%\left(
\kappa(\bm{x},t)
% \frac{k \mathcal{Q}(\bm{x},t)}{\mathcal{Q}_{max}(t)}
% \right) 
\mathbf{\Delta}\mathsfbi{C}(\bm{x},t).
\end{equation}
Notably, the simplified and homogeneous Laplacian form of the polymer artificial diffusivity term has been adopted in previous numerical studies utilizing the GAD scheme \cite{Alves_Rev, plan_gupta_vincenzi_gibbon_2017, gupta_vincenzi_2019} and the local artificial diffusivity (LAD) scheme \cite{Min_2001, Dubief_2005}. 
However, as mentioned in the main text, the authors note that $\kappa(\bm{x},t)$ [see, Eq.~(3) in main text] is one of several possible formulations for MAD. For instance, another possible alternative is to consider MAD in the form of the general diffusion equation in the case of variable diffusivity (denoted as MAD*), such that Eq.~\ref{eq:1_sup} can be presented as,  
\begin{equation}\label{eq:2_sup}
\mathsfbi{D}(\bm{x},t) = 
%\left(
\mathbf{\bnabla}\bcdot[\kappa(\bm{x},t)\mathbf{\bnabla}\mathsfbi{C}(\bm{x},t)],
\end{equation}
%whose form is akin to previous sub-grid scale models, such as large-eddy simulation (LES) \cite{LES_REF}.
whereby, the Laplacian form is instead replaced with a divergence operation. 

When compared to MAD for the CF case, the alternative MAD* scheme leads to completely different dynamics observed in the hydrodynamic and polymer fields, as illustrated in Fig.~\ref{figure_2_supp}. These differences are illustrated even in the the early stages of the flow [refer to Fig.~\ref{figure_2_supp} (a)]  where the rate of energy transfer from kinetic energy to elastic energy is slower for MAD*. After initially experiencing chaotic fluctuations, at the later stages, MAD* transitions to periodic dynamics. Qualitatively, during this periodic regime the polymer field for MAD* is characterised by double-leading vortices [refer to Fig.~\ref{figure_2_supp} (b)]. 
Despite this periodic regime resembling the results for GAD [refer to Fig.~3 (b) in the main text], we suspect this relaxation towards a periodic regime to be attributed to limited periodicity, as opposed to a continuous build-up of artificial diffusion (i.e., GAD), thus implying that long-range polymer interactions exceed half the characteristic unit cell \cite{vedad_jfm}.
The differences across the artificial diffusivity schemes is further exemplified with FRM in Fig.~\ref{figure_3_supp}.
More specifically, the time series of the dimensionless mean kinetic energy [refer to Fig.~\ref{figure_3_supp} (a)] shows that similar to GAD, MAD* experiences a dramatic transition into flow asymmetry, characterised by dominant vortices [refer to Fig.~\ref{figure_3_supp} (b)]. However, compared with GAD, with MAD* this initial transition occurs earlier and features of ET is recovered at a faster rate (i.e., flow symmetry and chaotic fluctuations), as illustrated in Fig.~\ref{figure_3_supp} (c). Despite this, at the later stages, MAD* fails to experience the same level of high-frequency oscillations in the hydrodynamic field, as observed for MAD. The initial chaotic regime observed for CF [Fig.~\ref{figure_2_supp} (a)] and the faster rate of recovery of ET features observed for FRM [Fig.~\ref{figure_3_supp} (c)], demonstrate that overall MAD* (\ref{eq:2_sup}) retains ET features better than GAD, however, does not offer performance comparable with MAD (\ref{eq:1_sup}). The source of discrepancies observed for MAD and MAD* is still currently unclear. Nevertheless, these results further highlight the complex interplay between between artificial diffusivity and periodic boundary conditions. 

\begin{figure}
	\centering
	\includegraphics[scale=0.75]{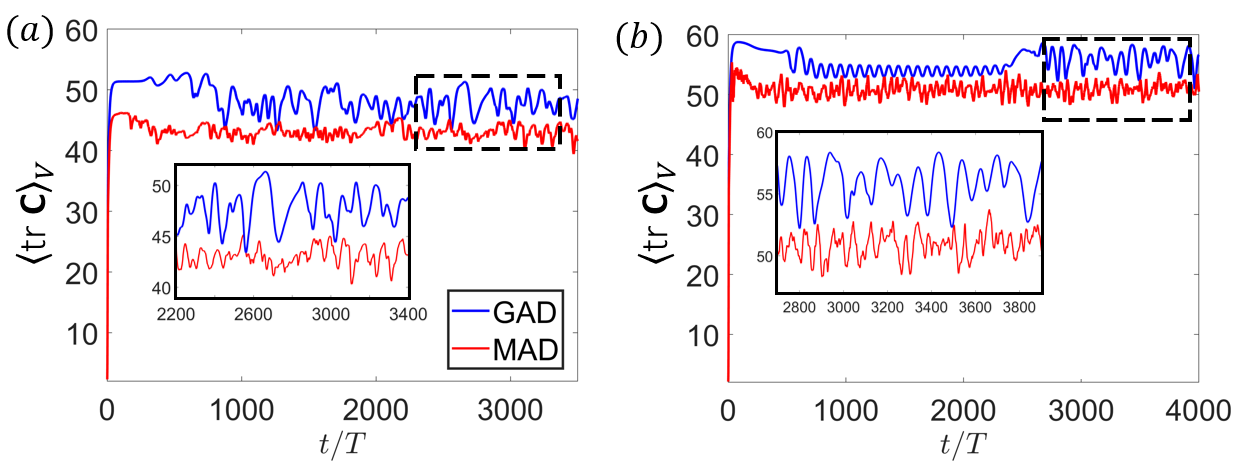}%width=0.95\textwidth
	\caption{Time series of the spatially averaged polymer trace $\langle $tr$(\mathsfbi{C})\rangle_{V}$ for the (a) FRM and (b) CF using the GAD (blue) and MAD (red) schemes. Note, the insets provided in (a) and (b) are zoom-in sections corresponding to the late-time dynamics in the dashed box regions. }
	\label{figure_1_supp}
\end{figure}
\section{Polymer field results}
In Fig.~\ref{figure_1_supp} the time history response for the spatially averaged polymer trace, $\langle $tr$(\mathsfbi{C})\rangle_{V}$, is compared for both artificial diffusivity schemes using FRM and CF. The results for the polymer field are analogous with the results obtained for the hydrodynamic field in Fig.~3 of the main text. Whereby, for both FRM and CF, GAD is shown to slowly transition into an unsteady regime over time, whereas MAD instantly transitions into the chaotic ET regime. Notably, the slow transition for GAD was also highlighted in the main text, and arises due to the excessive artificial diffusivity suppressing intrinsic instabilities. Furthermore, within the chaotic regime, even for the polymer field the fluctuations for MAD are at a much higher frequency compared to the slow aperiodic state observed for GAD, further supporting the results presented in the main text.

\section{Cellular Forcing Scheme Animations}
The animations provided in Movie1 and Movie2 correspond to the polymer trace field for CF using GAD and MAD, respectively. For GAD, the transition to an unsteady state is delayed, and results in an initial loss of forcing symmetry which fluctuates between two flow states. Over time, the excessive smearing of the polymer stress from GAD results in a complete loss of symmetry, characterised by two dominant vortices. Beyond this state, the polymer field attempts to recover the forcing symmetry, but is still largely affected by the strong presence of artificial diffusivity. On the other hand, the animation for MAD shows that the polymer field instantly transitions to a chaotic ET regime, where the heavy fluctuations perturb the cellular vortices, however, the large-scale structure remains largely constrained to the background forcing with momentary losses of symmetry.

\bibliography{BIB_ET_PAPER}